\documentclass[11pt]{article}
\usepackage{amssymb}
\newcommand{\equal}{\!\!\!&=&\!\!\!}

\begin{document}
\abovedisplayshortskip 12pt
\belowdisplayshortskip 12pt
\abovedisplayskip 12pt
\belowdisplayskip 12pt
\baselineskip=15pt
\title{{\bf Hamiltonian Structures for the\break Ostrovsky-Vakhnenko Equation}}
\author{J. C. Brunelli\thanks{\texttt{jcbrunelli@gmail.com}}  \\
\\
Departamento de F\'\i sica, CFM\\
Universidade Federal de Santa Catarina\\
Campus Universit\'{a}rio, Trindade, C.P. 476\\
CEP 88040-900\\
Florian\'{o}polis, SC, Brazil\\
\\
\\
S. Sakovich\thanks{\texttt{saks@tut.by}} \\
\\
Institute of Physics\\
National Academy of Sciences\\
220072 Minsk, Belarus\\
}
\date{}
\maketitle

\begin{center}
{ \bf Abstract}
\end{center}

We obtain a bi-Hamiltonian formulation for the Ostrovsky-Vakhnenko (OV) equation using its higher order symmetry
and a new transformation to the Caudrey-Dodd-Gibbon-Sawada-Kotera equation. Central to this derivation is the relation between Hamiltonian structures when dependent and independent variables are transformed.
\bigskip



\newpage

\section{Introduction:}

The Whitham equation
\begin{equation}
u_t + uu_x+\int_{-\infty}^{+\infty}K(x-y)u_ydy=0\label{whitham}
\end{equation}
was introduced in \cite{Whitham1974} to model a wave equation containing both breaking and peaking. A good discussion of equations (\ref{whitham}) can be found in \cite{Liu2006}. Here we are interested in equation (\ref{whitham}) with the kernel
\[
K(x-y)={1\over 2} |x-y|\;,
\]
which yields
\begin{equation}
u_t+uu_x+\partial^{-1}u=0\;,\label{VE}
\end{equation}
or in a local form
\begin{equation}
(u_t+uu_x)_x+u=0\;.\label{localVE}
\end{equation}
Equation (\ref{localVE}) also follows as a particular limit of the following generalized Korteweg-de Vries (KdV) equation
\begin{equation}
(u_t+uu_x-\beta u_{xxx})_x=\gamma u\;,\label{gkdv}
\end{equation}
derived by Ostrovsky \cite{Ostrovsky1978} to model small-amplitude long waves in a rotating fluid ($\gamma u$ is induced by the Coriolis force) of finite depth. For $\beta=0$ (no high-frequency dispersion)  the equation (\ref{gkdv}) is known under different names in the literature, such as the reduced Ostrovsky equation, the Ostrovsky-Hunter equation, the short-wave equation and the Vakhnenko equation. From now on we will call (\ref{localVE}) the Ostrovsky-Vakhnenko (OV) equation.
This OV equation describes the short-wave perturbation in a relaxing medium \cite{Vakhnenko1992}. This equation has a purely dispersive term and although it has the same nonlinearity of the KdV equation the dispersive terms are different. In \cite{Vakhnenko1992} and in a series of papers \cite{Parkes1993,VakhnenkoParkes1998,MorrisonParkesVakhnenko1999} it was established its integrability by deriving explicit solutions. Also, in \cite{ParkesVakhnenko2002} the integrability via inverse scattering method was derived via a third-order eigenvalue problem obtained after a B{\"{a}}cklund transformation. However in \cite{HoneWang2003} this Lax pair was written in its original variables as a zero curvature condition
\[
\mathbb{A}_{1,t}-\mathbb{A}_{0,x}-[\mathbb{A}_0,\mathbb{A}_1]=0\;,
\]
with
\[
\mathbb{A}_1={1\over3}
\left(
\begin{array}{ccc}
0 &  u_x/\lambda &-1/\lambda\\
\noalign{\vskip 12pt}
-1 &  0 &0\\
\noalign{\vskip 12pt}
-u_x &  -1 &0
\end{array}
\right)\qquad
\mathbb{A}_0={1\over3}
\left(
\begin{array}{ccc}
0 &  3-uu_x/\lambda &u/\lambda\\
\noalign{\vskip 12pt}
u &  0 &3\\
\noalign{\vskip 12pt}
3\lambda+uu_x &  -2u &0
\end{array}
\right)\;.
\label{A1A0}
\]
Also,  they obtained the following third order Lax pair for the OV equation
\begin{eqnarray}
\psi_{xxx}+{1\over 9\lambda}\left(u_{xx}+{1\over 3}\right)\Psi\equal 0\;,\nonumber\\\noalign{\vskip 7pt}
\psi_t-9\lambda\psi_{xx}+u\psi_x-u_x\psi\equal 0\;,\label{lax}
\end{eqnarray}
whose compatibility condition yields the $x$-derivative of the OV equation (\ref{localVE}).

It turns out that the OV equation can be obtained as a short-wave limit of another integrable equation, the Degasperis-Procesi (DP) equation \cite{DegasperisProcesi1999,DegasperisHolmHone2002,DegasperisHolmHone2003}
\begin{equation}
u_t-u_{xxt}+4uu_x=3u_xu_{xx}+uu_{xxx}\;.\label{DPE}
\end{equation}
The authors of \cite{DegasperisHolmHone2002,DegasperisHolmHone2003} found a Lax pair, derived two infinite sequences of conserved charges for (\ref{DPE}) and proposed a bi-Hamiltonian formulation. Hone and Wang \cite{HoneWang2003} have shown and explored the fact that the OV equation can be obtained as a limit of the DP equation through the transformation
\[
T: \left\{ \begin{array}{l}
x\to\displaystyle{\epsilon x-{t\over3\epsilon}}\\
t\to\displaystyle{\epsilon t}\\
u\to\displaystyle{u-{1\over 3}\epsilon^2}
\end{array} \right.\label{T}
\]
in the short wave limit $\epsilon\to 0$,
\[
{\rm DP}\,\,{\buildrel T\over\longrightarrow}\,\,{\rm OV}-\epsilon^2(u_t+4uu_x)\;.
\]
Using this short wave limit Hone and Wang obtained the scalar linear problem (\ref{lax}) from the corresponding Lax pair of the DP equation. Also, from the bi-Hamiltonian study of the DP equation performed in \cite{DegasperisHolmHone2002} they proposed two Hamiltonian operators for the OV equation but not in the original variable $u$.

In this paper our main interest is to investigate the Hamiltonian integrability for the OV equation directly in its original variable and evolutionary nonlocal form (\ref{VE}). As far as we known this is the simplest equation involving nonlinearity an nonlocality and will provide a good ``laboratory" for study of nonlocal equation.

Results on the OV equation frequently rely on the well known reciprocal transformation to the special case of Ito's equation  which can be written in Hirota's bilinear form \cite{VakhnenkoParkes1998}. We point out that a transformation between the OV equation and the Bullough-Dodd-Tzitzeica equation is also known \cite{MannaNeveu2001} but not explored in the literature. In this paper we will introduce and explore a new third transformation. In Section \ref{sectionsymmetry} we find a fifth order symmetry to the OV equation and we show in Section \ref{sectionCDGSK} that the evolution equation associated to this symmetry can be transformed to the Caudrey-Dodd-Gibbon-Sawada-Kotera equation (CDGSK) \cite{SawadaKotera1974,Caudrey,Fuchssteiner1982} via a chain of Miura-type transformations. In Section \ref{sectionsymmetry} we also find the first Hamiltonian structure for the OV hierarchy. The transformations introduced in Section \ref{sectionCDGSK} transform dependent as well as independent variables and we give in Section \ref{sectionchangeofvariables} a formula relating the Hamiltonian structures in this situation. This formula is used in Section \ref{sectionsecondhamiltonian} to obtain the second Hamiltonian structure of the OV equation via the well known CDGSK's Hamiltonian structure. Our results are purely algebraic and no analytical justifications are provided. Questions about wave breaking, blow-up rates, boundary conditions, well-posedness and so on can be found, for instance, in \cite{LiuPelinovskySakovich2010} and references within.

\section{Symmetries and a first Hamiltonian structure}\label{sectionsymmetry}

First let us look for some symmetries of the OV equation (\ref{VE}). It is well known \cite{Olver1993,Blaszak1998} that a symmetry $\sigma$ of the evolution equation $u_t=K(u)$ should satisfy
\begin{equation}
{\partial\sigma\over\partial t}+[\sigma,K]=0\;,\label{symmetrycond}
\end{equation}
where
\[
[\sigma,K](u)\equiv \sigma'(u)[K(u)]-K'(u)[\sigma(u)]\label{commutator}
\]
and
\begin{equation}
F'(u)[X]={d\ \over d\epsilon}F(u+\epsilon X)\Big|_{\epsilon=0}\label{frechet}
\end{equation}
is the Fr\'{e}chet or directional derivative of $F(u)$ in the direction of the vector field $X$. In the case that $F(u,v)$ depends on two variables we write $F_u$ and $F_v$ for the Fr\'{e}chet derivative with respect to $u$ and $v$, respectively. If $H$ is a functional then
\begin{equation}
H'(u)[X]=\langle {\rm grad\,}H,X\rangle=\left\langle {\delta H\over\delta u},X\right\rangle\;,\label{grad}
\end{equation}
where $\langle\cdot,\cdot\rangle$ denotes the usual scalar product between the dual spaces. In what follows, we indicate by ``$*$" the transpose of a operator with respect to the duality. From (\ref{symmetrycond}) it is easy to check that
\begin{equation}
\sigma_0= 0\;,\quad
\sigma_1= K(u)\;,\quad
\sigma_2= u_x\;,\quad
\sigma_3= -{1\over 2}\left[(1+3u_{xx})^{-2/3}\right]_{xxx}\label{symmetries}
\end{equation}
are symmetries for the autonomous OV equation (\ref{VE}). This means that the OV equation and the local nonlinear equation
\begin{equation}
u_t=-{3\over 2}D_x^3\left(u_{xx}+{1\over3}\right)^{-2/3}\;,\label{newequation}
\end{equation}
where the right-hand side is the generator $\sigma_3$ of the higher symmetries and where the coefficient $-3/2$ is taken for simplicity, are members of the same hierarchy of equations and share many properties such as integrability. As far as we know this Ostrovsky-Vakhnenko fifth-order equation (OVF) (\ref{newequation}) is a new integrable equation but we will show that it can be transformed to the CDGSK equation in the next Section.

By construction the OV equation (\ref{VE}) and the OVF equation (\ref{newequation}) share the same conserved charges as well Hamiltonian structures. Let us find the first Hamiltonian structure for these equations \cite{Brunelli2005}. Introducing the Clebsch potential $u=\phi_x$ the equation (\ref{VE}) can be written as
\[
\phi_{xt}+\phi_x\phi_{xx}+\phi=0\;.
\]
This equation can be obtained from a variational principle,
$\delta\int dtdx\,{\cal L}$, with the Lagrangian density
\begin{equation}
{\cal L}={1\over 2}\phi_t\phi_x+{1\over 6}\phi_x^3-{1\over 2}\phi^2\;.\label{lagrangian}
\end{equation}
This is a first order Lagrangian density and we can use, for
example, Dirac's theory of constraints to obtain the
Hamiltonian and the Hamiltonian operator associated with
(\ref{lagrangian}). The Lagrangian is degenerate and the primary
constraint is obtained to be
\begin{equation}
\Phi=\pi-{1\over 2}\phi_x\;,\label{primary}
\end{equation}
where $\pi={{\partial{\cal L}}/{\partial \phi_t}}$ is the canonical
momentum. The total Hamiltonian can be written as
\begin{equation}
H_T =\int dx\left(\pi \phi_t-{\cal L}+\lambda\Phi\right)
=\int dx\left[- {1\over6}\phi_x^3 + {1\over 2}\phi^2 +\lambda\left(\pi-{1\over
2}\phi_x\right)\right]\;,\label{ht}
\end{equation}
where $\lambda$ is a Lagrange multiplier field. Using the canonical
Poisson bracket relation
\begin{equation}
\{\phi(x),\pi(y)\}=\delta(x-y)\;,\label{poisson}
\end{equation}
with all others vanishing, it follows that the requirement of the
primary constraint to be stationary under time evolution, $\{\Phi(x),H_T\}=0$,
determines the Lagrange multiplier field $\lambda$ in (\ref{ht}) and
the system has no further constraints. Using the canonical Poisson bracket relations (\ref{poisson}), we
can now calculate
\begin{equation}
K(x,y)\equiv\{\Phi(x),\Phi(y)\}={1\over
2}D_y\delta(y-x)-{1\over
2}D_x\delta(x-y)\;.\label{kpoisson}
\end{equation}
This shows that the constraint (\ref{primary}) is second class and
that the Dirac bracket between the basic variables has the form
\[
\{\phi(x),\phi(y)\}_D=\{\phi(x),\phi(y)\}-\int
dz\,dz'\{\phi(x),\Phi(z)\}J(z,z')\{\Phi(z'),\phi(y)\}=J(x,y)\;,\
\]
where $J$ is the inverse of the Poisson bracket of the constraint
(\ref{kpoisson}),
\[
\int dz\,K(x,z) J(z,y)=\delta(x-y)\,.
\]
This last relation determines $D_x J(x,y)=-\delta(x-y)$ or $J(x,y)={\cal D}\delta(x-y)$
where
\begin{equation}
{\cal D}=-D_x^{-1}\;.\label{d}
\end{equation}
We can now set the constraint (\ref{primary})
strongly to zero in (\ref{ht}) to obtain
\[
H_T =\int dx\left(-{1\over6}\phi_x^3+{1\over
2}\phi^2\right)\;.
\]
Using (\ref{d}) and the transformation properties of Hamiltonian operators (see (\ref{dtransformed})), we get
\[
{\cal D}=D_x\left({\cal D}\right)(D_x)^*=D_x\;,
\]
and the OV equation (\ref{VE}) can be written in the Hamiltonian form as
\begin{equation}
u_t={\cal D}_1{\delta H_1\over\delta u}\;,\qquad {\cal D}_1=D_x\;,\qquad H_1=\int dx\left[-{1\over 6}u^3+{1\over 2}(\partial^{-1}u)^2\right]\;.\label{firsths}
\end{equation}
It can be easily checked that $H_1$ is conserved by both the OV and OVF equations for rapidly decreasing or periodic boundary conditions.  Taking the second $x$ derivative of the OV equation we obtain the trivial conserved charge
\begin{equation}
\int dx\, u_{xx}\;,\label{htrivial}
\end{equation}
and from this equation times $(u_{xx}+{1/3})^{-2/3}$ we also get
\[
H_2=-{9\over 2}\int dx \left(u_{xx}+{1\over 3}\right)^{1/3}\;,
\]
as conserved charge for both the OV and OVF equations. In this way the OVF equation has the following Hamiltonian representation
\[
u_t={\cal D}_1{\delta H_2\over\delta u}\;,\qquad {\cal D}_1=D_x\;,\qquad H_2=-{9\over 2}\int dx \left(u_{xx}+{1\over 3}\right)^{1/3}\;.
\]
Multiplying the OV equation (\ref{VE}) by $u$ and integrating we also get the following conserved charge
\begin{equation}
\int dx\, u^2\;.
\end{equation}

\section{Transformation to the CDGSK equation}\label{sectionCDGSK}

The link between the KdV and modified KdV equations through a Miura transformation is not an isolated result in the theory of integrable models. Very often a nonlinear equation is equivalent to a known and well studied equation  via a chain of Miura-type transformations and now we will show that we can relate the OVF equation with the CDGSK equation using this procedure. We follow the methods used in \cite{Sakovich1991,Sakovich1993,Sakovich2003,Sakovich2005}.

The transformation
\begin{equation}
(x,t,u(x,t))\mapsto (x,t,v(x,t)): \begin{array}{l}
v=(u_{xx}+1/3)^{-1/3}
\end{array} \label{T1}
\end{equation}
relates the OVF equation with
\begin{equation}
v_t={1\over 2}v^4D_x^5 v^2\;.\label{fujimoto}
\end{equation}
The right-hand side of the transformation (\ref{T1}) is obtained from the separant of the OV equation. (The separant of an evolution equation $u_t=f(u,u_x,\dots,u_{(n)})$ is $\partial f/\partial u_{(n)}$, where $u_{(n)}$ is the highest derivative of $u$ with respect to $x$.) Let us note that (\ref{fujimoto}) is one of the non-constant separant evolution Fujimito-Watanabe equations (see \cite{Sakovich1991} and references within). The equation (\ref{fujimoto}) has a separant $v^5$ and by the Ibragimov substitution
\begin{equation}
(y,t,w(y,t))\mapsto (x,t,v(x,t)):\left\{ \begin{array}{l}
v(x,t)=w_y(y,t)\\
x=w(y,t)
\end{array} \right.\label{T2}
\end{equation}
is related with the constant separant equation
\begin{equation}
w_t=w_{5y}-5w_y^{-1}w_{2y}w_{4y}+5w_y^{-2}w_{2y}^2w_{3y}\;,\label{wequation}
\end{equation}
where $w_{ky}=\partial_y^kw$, $k=2,3,4,5$ (and similar notations for derivatives are used in what follows). Finally, the transformation
\begin{equation}
(y,t,w(y,t))\mapsto (y,t,z(x,t)):\begin{array}{l}
z=-w_y^{-1}w_{3y}
\end{array} \label{T3}
\end{equation}
relates (\ref{wequation}) with the CDGSK equation
\begin{equation}
z_t=z_{5y}+5zz_{3y}+5z_yz_{2y}+5z^2z_y\;.\label{CDGSKE}
\end{equation}
In summary, we have schematically
\[
\begin{tabular}{ccccccc}
 &$B(u,v)=0$ && $B(w,v)=0$ && $B(w,z)=0$ &\\
\noalign{\vspace{-.1truecm}}
 OVF & $\longrightarrow$ & (\ref{fujimoto}) & $\longleftarrow$ & (\ref{wequation}) & $\longrightarrow$ & CDGSK,
\end{tabular}
\]
where the implicit transformations are
\begin{eqnarray}
&&B(w,z)(y)=\left(z+w_y^{-1}w_{3y}\right)(y)\;,\nonumber\\
&&B(w,v)(y)=\left(v(w)-w_y\right)(y)\;,\nonumber\\
&&B(u,v)(x)=\left(v-(u_{xx}+1/3)^{-1/3}\right)(x)\;.\label{BBB}
\end{eqnarray}

\section{Hamiltonian structure behavior under change of variables}\label{sectionchangeofvariables}

Transformations between two evolution equations generate transformations between the corresponding structures such as recursion operators, Hamiltonian structures, conserved charges and so on \cite{Fokas1980,Fokas1981}. Let be the change of variables among dependent and independent variables
\begin{equation}
\left\{ \begin{array}{l}
y=P(x,u^{(n)})\;,\\
\noalign{\vskip 5pt}
v=Q(x,u^{(n)})\;,
\end{array} \right.\label{transformxv}
\end{equation}
where $u^{(n)}$ represents all derivatives of $u$ with respect to $x$ of order at most $n$. We want to relate the Hamiltonian representations
\begin{eqnarray}
&&u_t={\cal D}^{(u)}{\delta H\over\delta u}={\cal D}^{(u)}{\rm E}_u(h)\;,\label{equ}\\
\noalign{\vskip 1pt}
&&{v}_t=\widetilde{\cal D}^{(v)}{\delta \widetilde{H}\over\delta v}=\widetilde{\cal D}^{(v)}{\rm E}_{v}(\tilde{h})\label{eqv}\;,
\end{eqnarray}
where
\begin{eqnarray}
&&H[u]=\int dx\,h(x,u^{(n)})\;,\label{hamu}\\
\noalign{\vskip 1pt}
&&\widetilde{H}[v]=\int dy\,\tilde{h}(y,{v}^{(n)})\;,\label{hamv}
\end{eqnarray}
and ${\rm E}_u(h)$ is the Euler operator  acting on the Hamiltonian density $h$.
The transformation (\ref{transformxv}) defines an implicit function $B(u,v)=0$, and we have
\[
B_uu_t+B_vv_t=0\;,
\]
or
\begin{equation}
v_t=-Tu_t\;,\label{vtut}
\end{equation}
where
\begin{equation}
T=B_v^{-1}B_u\;,
\end{equation}
and $B_u$ and $B_v$ are the Fr\'{e}chet derivatives defined in (\ref{frechet}). Now
we use the following result (see \cite{Olver1993}, Exercise 5.49, pg. 386) for the relation between the action of the Euler operator under a change of variables
\begin{equation}
{\rm E}_u(h)={\cal O}\,{\rm E}_{v}(\tilde{h})\;,\label{euev}
\end{equation}
where
\begin{equation}
{\cal O}(R)=Q_u^*(D_xP\cdot R)-P_u^*(D_xQ\cdot R)\;.\label{operatoro}
\end{equation}
From (\ref{vtut}), (\ref{equ}) and (\ref{euev})
\[
v_t=-T{\cal D}^{(u)}{\cal O}\,{\rm E}_{v}(\tilde{h})
\]
and comparing with (\ref{eqv}) we finally get
\begin{equation}
\widetilde{\cal D}^{(v)}=-T{\cal D}^{(u)}{\cal O}\;.\label{dtransformed}
\end{equation}
When the independent variable is not transformed $B_u=-Q_u$, $B_v=1$, $T=-Q_u$ and ${\cal O}=Q_u^*$. Also, when we calculate how a recursion operator $R={\cal D}_2{\cal D}_1^{-1}$ transforms under (\ref{transformxv}) the operator ${\cal O}$ drops out. These are the usual results commonly found in the literature \cite{Fokas1980,Fokas1981}. The generalization of (\ref{dtransformed}) for a number of  dependent and independent variables greater than one is straightforward.

\section{Second Hamiltonian structure}\label{sectionsecondhamiltonian}

We use the results of the last Section to obtain the second Hamiltonian structure of the OV equation (\ref{VE}) from the known Hamiltonian structure of the CDGSK equation (\ref{CDGSKE}) given by \cite{Fuchssteiner1982}

\begin{equation}
{\cal D}^{(z)}=D_y^3+2(zD_y+D_yz)\;,\qquad H^{(z)}=\int dy\left({1\over 6}z^3-{1\over2}z_y^2\right)\;.\label{CDGSKhamstruc}
\end{equation}

From (\ref{BBB}) and (\ref{vtut}) we have
\begin{eqnarray}
z_t=Aw_t\;,\quad&& A=-w_y^{-1}D_y^3+w_y^{-2}w_{3y}D_y=-v^{-1}D_xv^3D_x^2\;,\label{TA}\\
v_t=Bw_t\;,\quad&& B=D_y-w_y^{-1}w_{2y}=v^2D_xv^{-1}\;,\label{TB}\\
v_t=Cu_t\;,\quad&& C=-{1\over 3}\left(u_{xx}+{1\over 3}\right)^{-4/3}\!\!\!D_x^2=-{1\over 3}v^4D_x^2\;.\label{TC}
\end{eqnarray}
From (\ref{TA})--(\ref{TC}) we already see that is very convenient to use the variable $v(x,t)$ because this factorizes all the operators involved, via the relations $D_y=vD_x$, $w_y=v$, $w_{2y}=vv_x$, $w_{3y}=v(vv_x)_x$, $z=-(vv_x)_x$ and $u_{xx}=v^{-3}-1/3$; note also that
\begin{eqnarray}
&&D_y^2+z=D_xv^3D_xv^{-1}\;,\nonumber\\
\noalign{\vskip 5pt}
&&D_y^3+4zD_y+2z_y=v^{-1}D_xv^3D_xv^3D_xv^{-2}\;.\label{CDGSKv}
\end{eqnarray}
Now using (\ref{euev}) and (\ref{operatoro}) we have
\begin{eqnarray}
{\delta H^{(w)}\over\delta w}=\bar{A}{\delta H^{(z)}\over\delta z}\;,\quad&& \bar{A}={A}^*=D_y^3w_y^{-1}-D_yw_y^{-2}w_{3y}=vD_x^2v^3D_xv^{-2}\;,\label{OA}\\
{\delta H^{(w)}\over\delta w}=\bar{B}{\delta H^{(v)}\over\delta v}\;,\quad&& \bar{B}=-D_yw_y-w_{2y}=-D_xv^2\;,\label{OB}\\
{\delta H^{(u)}\over\delta u}=\bar{C}{\delta H^{(v)}\over\delta v}\;,\quad&& \bar{C}={C}^*=-{1\over 3}D_x^2\left(u_{xx}+{1\over 3}\right)^{-4/3}\!\!\!=-{1\over 3}D_x^2v^4\;.\label{OC}
\end{eqnarray}
Therefore, the Hamiltonian operators transform as
\begin{equation}
{\cal D}^{(w)}=A^{-1}{\cal D}^{(z)}\bar{A}^{-1}\;,\quad
{\cal D}^{(v)}=B{\cal D}^{(w)}\bar{B}\;,\quad
{\cal D}^{(u)}=C^{-1}{\cal D}^{(v)}\bar{C}^{-1}\;,
\end{equation}
and as result of (\ref{CDGSKv}), (\ref{TA})--(\ref{TC}) and (\ref{OA})--(\ref{OC}) we obtain the operator of order minus five
\begin{equation}
{\cal D}^{(u)}=9D_x^{-2}v^{-2}D_xv^{-1}D_x^{-3}v^{-1}D_xv^{-2}D_x^{-2}\label{newdu}\;,
\end{equation}
where now $v$ is not a dependent variable but just a placeholder for the expression in (\ref{T1}), i.e., $v\equiv (u_{xx}+1/3)^{-1/3}$.

Now, let us find the Hamiltonian $H^{(u)}$ corresponding to $H^{(z)}$ in (\ref{CDGSKhamstruc}). Its variational derivative is
\[
{\delta H^{(z)}\over\delta z}=z_{2y}+{1\over 2}z^2\;,
\]
and from (\ref{OA}) and (\ref{OB}) we can transform it into
\begin{eqnarray*}
&&{\hskip 3truecm}{\delta H^{(v)}\over\delta v}=\bar{B}^{-1}\bar{A}{\delta H^{(z)}\over\delta z}=\nonumber \\
&&=v^3 v_{6x} + 9 v^2 v_x v_{5x} + 17 v^2v_{2x} v_{4x} + 14 vv_x^2 v_{4x} + \frac{19}{2} v^2 v_{3x}^2+\nonumber\\
&& + 34 vv_x v_{2x} v_{3x} - 2 v_x^3 v_{3x} + \frac{20}{3} v v_{2x}^3- 10 v_x^2 v_{2x}^2 + 5 v^{-1}v_x^4 v_{2x} - {5\over6}v^{-2}{v_x^6}\;,
\end{eqnarray*}
and we can reconstruct the corresponding conserved charge using the homotopy formula
\[
H^{(v)}=\int dx\int_0^1d\lambda\,\, v{\delta H^{(v)}(\lambda v)\over \delta v}\!\!\;,
\]
to finally obtain
\begin{equation}
H^{(v)}=\int dx\left( - \frac{1}{2} v^3 v_{3x}^2 + \frac{4}{3} v^2 v_{2x}^3 - 2 v v_x^2 v_{2x}^2 - {1\over 6}v^{-1}v_x^6\right)\;,\label{newhv}
\end{equation}
which is also $H^{(u)}$ if we make the substitution (\ref{T1}) to the right hand side of (\ref{newhv}). Computationally, this is easier than to directly perform the calculations in the $u$ variable.

From (\ref{newdu}) and (\ref{newhv}) we obtain the following second Hamiltonian structure for the OVF equation (\ref{newequation})
\begin{eqnarray*}
&&u_t={\cal D}_2{\delta H_3\over\delta u}\;,\nonumber\\
\noalign{\vskip 5pt}
&&{\cal D}_2=9D_x^{-2}v^{-2}D_xv^{-1}D_x^{-3}v^{-1}D_xv^{-2}D_x^{-2}\;,\nonumber\\
\noalign{\vskip 5pt}
 && H_3=-{1\over2}\int dx\left( - v^3 v_{3x}^2 - \frac{8}{3} v^2 v_{2x}^3 +4 v v_x^2 v_{2x}^2 +\frac{1}{3} v^{-1} v_x^6\right)\;,\label{}
\end{eqnarray*}
where $v$ is given by (\ref{T1}). However, the second Hamiltonian formulation for the OV equation (\ref{VE}) can be written formally as
\begin{equation}
u_t={\cal D}_2{\delta H_4\over\delta u}\;,\qquad {\cal D}_2=9D_x^{-2}v^{-2}D_xv^{-1}D_x^{-3}v^{-1}D_xv^{-2}D_x^{-2}\;,\qquad H_4=-{1\over2}\int dx\, u_{xx}\;,\label{secondhs}
\end{equation}
where we have used the trivial conserved charge (\ref{htrivial}) with $-2\delta H_4/\delta u=D_x^2\cdot1$ and $D_x^{-2}D_x^2\cdot1=1$. This same charge and behavior appears in the pull-back of the Harry Dym equation to the Hunter-Saxton equation \cite{HunterZheng} and is due to the nonlocality of the Hamiltonian operator ${\cal D}_2$.

From the CDGSK conserved charge $H^{(z)}=\int dy\,z$ we get
\[
\int dx \,{v^{-1}v_x^2}\;,
\]
for the OV equation. In fact, from (\ref{firsths}) and (\ref{secondhs}) we have the recursion operator
\[
R={\cal D}_2{\cal D}_1^{-1}=9D_x^{-2}v^{-2}D_xv^{-1}D_x^{-3}v^{-1}D_xv^{-2}D_x^{-3}\;,
\]
and this could be the starting point to generate a hierarchy of equations and charges for the OV system of equations which we will explore in a future publication.

\end{document}